\documentclass[twocolumn, a4paper, 10pt]{article}
\usepackage{epsfig}

\begin{document}

\title {\normalsize \bf SEARCHING 
FOR GRAVITATIONAL LENSES IN THE DISTANT UNIVERSE}
\author {\normalsize \bf Andrew W. Blain \\ \\
\normalsize Cavendish Laboratory, Madingley Road,
Cambridge, CB3  0HE, UK.\\
\normalsize E-mail: awb@mrao.cam.ac.uk; Tel: +44 1223 337234; Fax: +44 1223 
354599.
} 
\date{ }
\maketitle

\thispagestyle{empty} 

\vskip 0.5cm
{\centerline {ABSTRACT} }
\vskip 0.2cm

{\it FIRST} has the ideal combination of sensitivity and angular resolving power to 
determine the bright counts of galaxies over a large area of sky in the
submillimetre (submm) waveband. Important information about the evolution of 
galaxies at moderate redshifts will be provided by the analysis of these counts,
and the effects of gravitational lensing will make the counts even more significant 
for observational cosmology. This paper describes the flux and surface densities 
of lensed galaxies in the submm waveband and explains how {\it FIRST} could 
be used to discover several tens of lenses. The {\it Planck} and {\it FIRST} 
missions are very well suited to cooperate in this programme, and together could 
compile a larger sample of several hundred lenses. {\it FIRST} will be able to 
detect the signal due to the Sunyaev--Zel'dovich (SZ) effect in a large sample of 
clusters. These observations could also be used to investigate submm-wave 
lensing by clusters, and to infer the properties of the population of faint distant 
galaxies. 

\vskip 0.5cm
{\centerline {1. INTRODUCTION} }
\vskip 0.2cm 

The submm-wave flux density of a distant ($z>1$) dusty star-forming galaxy or 
quasar with a fixed luminosity is predicted to be almost independent of redshift 
[Ref.\,1,\,2]. The continuing development of submm-wave cosmology is motivated 
by this flux density--redshift relation and the corresponding steep counts.
However, there is another counter-intuitive feature of the population of distant 
galaxies in the submm waveband -- the fraction of gravitationally-lensed 
galaxies is expected to be up to three orders of magnitude larger in a sample of 
galaxies that is selected in this waveband as compared with one selected in 
any other waveband [Ref.\,3,\,4]. The largest excess of lenses is expected at the 
flux density at which the submm-wave counts begin to rise steeply above the 
Euclidean slope. Gravitational lensing leads to a similar enhancement of the 
surface and flux densities of distant galaxies in the fields of clusters, which can 
be exploited to investigate the evolution of the population of faint background 
galaxies in submm-wave observations of the SZ effect [Ref.\,5]. A density 
parameter $\Omega_0=1$, a cosmological constant $\Lambda_0=0$ and a value 
of Hubble's constant $H_0=50$\,km\,s$^{-1}$\,Mpc$^{-1}$ are assumed 
throughout unless otherwise stated.  

\vskip 0.5cm
{\centerline {2. GRAVITATIONAL LENSING} }
\vskip 0.2cm

The properties of both the population and individual appearance of distant 
galaxies can be modified significantly by the gravitational lensing effect of 
foreground masses [Ref.\,6], whether galaxies, clusters or structures on larger
scales [Ref.\,7]. Gravitational lensing has been studied in most detail in the 
optical and radio wavebands, in which faint distant galaxies can be observed with 
sub-arcsecond resolution [Ref.\,8,\,9]. However, there are excellent prospects for
extending these studies into the submillimetre waveband, using both 
high-resolution and wide-field observations. 

The concept of magnification bias provides a useful way of describing the effects 
of lensing on a population of distant galaxies [Ref.\,8,\,10]. If the surface density 
of galaxies with flux densities larger than $S_\nu$ per unit redshift at redshift $z$ 
is $n(S_\nu, z)$, then magnification by a factor $A$ due to gravitational
lensing would predict a modified count, 
\begin{equation}
n'(S_\nu, z) = A^{-1} \, n[S_\nu A^{-1}, z]. 
\end{equation}
If $n \propto S_\nu^\alpha$ locally, then 
$n' \propto S_\nu^\alpha A^{-(1+\alpha)}$. The magnification bias $n'/n$ is hence 
given by $A^{-(1+\alpha)}$: for $\alpha < -1$ the surface density of galaxies is 
increased and the magnification bias is positive; for $\alpha > -1$ the surface 
density of galaxies is decreased and the magnification bias is negative. 
Magnification bias was first discussed in the context of samples of bright 
quasars; however, counts at faint flux densities in the submm waveband are 
uniquely steep, with $\alpha = -3$ or less, and so the magnification bias for faint 
galaxies in this waveband is expected to be very significant [Ref.\,4,\,11]. In the 
case of lensing by clusters of galaxies, the magnification $A$ is a function of 
position within the cluster, but the effect of magnification bias still applies. 

The magnification distribution as a function of redshift, $F(A,z)$ can be derived 
from the mass distribution of galaxies [Ref.\,12,\,13], and takes the form 
$a(z) A^{-3}$ if $A$ is large. A range of estimates of $a(z)$, which are calculated 
in several different world models and both evolving and non-evolving models of 
the distribution of lensing masses, are presented in Fig.\,1. The world models are 
all flat, with $\Omega_0 + \Lambda_0 = 1$, but include different density
parameters $\Omega_0$. The evolving mass distribution of the population of 
lensing galaxies is derived using the Press--Schechter formalism for structure 
formation by hierarchical clustering [Ref.\,14], in which galaxies typically become 
smaller and more numerous as redshift increases. The probability of lensing is 
expected to increase as the size of the cosmological constant increases, and is
expected to be smaller in the evolving model as compared with the
non-evolving model. The magnification distribution can be used to predict the 
surface density of lensed galaxies with flux densities greater than $S_\nu$,
\begin{equation}
N_{\rm L}(S_\nu) = \int_0^{z_0} \int_2^\infty {F(A,z) \over A}\,
n\left[{ {S_\nu} \over A}, z \right]\,{\rm d}A\,{\rm d}z.  
\end{equation}
The count of unlensed galaxies, 
\begin{equation}
N_{\rm U}(S_\nu) = \int_0^{z_0} n(S_\nu,z)\,{\rm d}z.
\end{equation}

\begin{figure}
\begin{center}
\epsfig{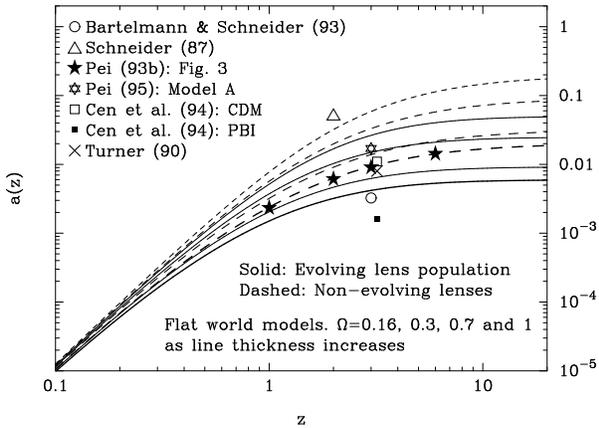}
\end{center}
\caption{The probability of galaxy--galaxy lensing as a function of redshift, 
which can change significantly if $\Lambda \ne 0$ or the lens population evolves. 
The lensing optical depth $\tau \simeq a/8$. For references see [Ref.\,4].}
\end{figure} 

\vskip 0.5cm
{\centerline {3. LENSING BY FIELD GALAXIES} }
\vskip 0.2cm

Models of the population of distant dusty star-forming galaxies [Ref.\,15] can be 
used to predict~$n$ (Equation\,1), and hence to derive counts of both lensed and
unlensed galaxies (Equations\,2\,\&\,3). The form of $a(z)$ assumed in
the calculation of the lensed counts is normalised to match the predictions of 
[Ref.\,13], that include a total density in compact objects $\Omega_{\rm G}=0.16$.

Multi-waveband studies of the evolution of galaxies and active galactic nuclei 
(AGN) [Ref.\,16,\,17,\,18\,19] indicate that the increase in both the global 
star-formation rate and the luminosity density of AGN is consistent with pure 
luminosity evolution of the form $(1+z)^3$ out to $z \sim 2$. Three models of 
galaxy evolution are discussed here; A, B and C. They are all normalised to match 
the low-redshift luminosity function of {\it IRAS} galaxies [Ref.\,20] and undergo 
pure $(1+z)^3$ luminosity evolution out to a redshift $z_{\rm max}=2$, but 
include different forms of evolution at larger redshifts $z < z_0$. Models~A and B
include no further luminosity evolution at $z>z_{\rm max}$ and values of 
$z_0 = 5$ and 2 respectively. Model~C includes negative luminosity evolution of 
the form $(1+z_{\rm max})^3\, t(z) / t(z_{\rm max})$ for 
$z_{\rm max}<z\le z_0(=10)$; $t(z')$ is the cosmic epoch at redshift $z'$. 

The counts predicted in all three models are compared in Fig.\,2(a). Estimates of 
the lensed counts are calculated for both the evolving and non-evolving mass 
distributions of lenses; the surface density of lensed images is predicted to be 
smaller in the evolving model. The counts predicted in different world models 
(Fig.\,1) are compared in Fig.\,2(b). A non-zero cosmological constant both 
increases the predicted surface density of lensed galaxies, due to an increased 
probability of lensing, and decreases the counts of unlensed galaxies, due to a 
smaller volume element at large redshifts as compared with an Einstein--de 
Sitter model. The wavelength dependence of the counts is presented in 
Figs\,2(c)\,\&\,2(d). The 90-$\mu$m counts take a similar form to those expected 
in the optical or radio wavebands, but the relative density of lensed and 
unlensed galaxies in the submm waveband is expected to be about two orders of 
magnitude larger as compared with other wavebands. The surface density of
both lensed and unlensed galaxies is expected to decrease as the wavelength
of observation increases.

\begin{figure*}[t]
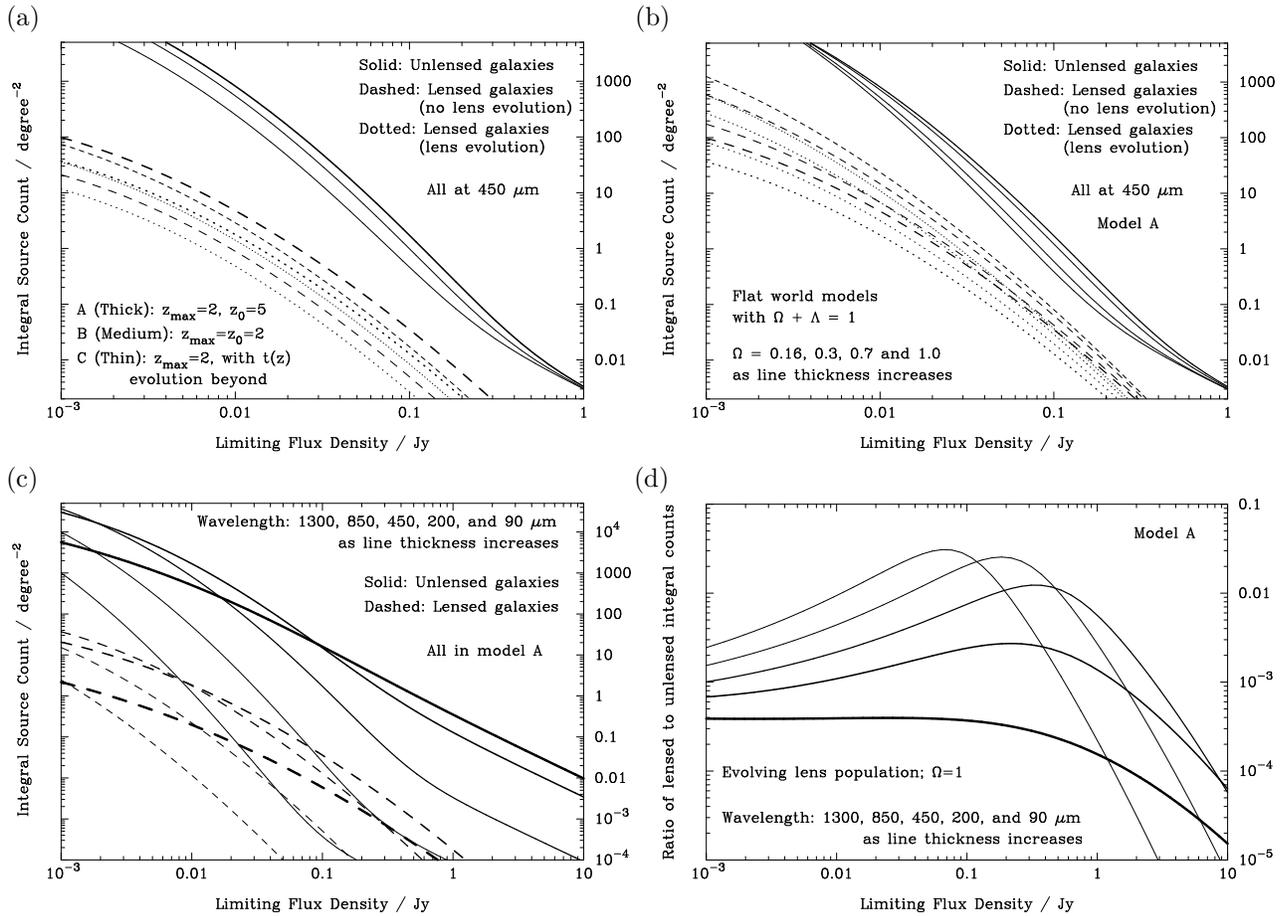

\begin{minipage}{170mm}
(a) \hskip 78mm (b)
\begin{center}
\epsfig{file=Gren_evmods.ps, 
width=4.95cm,bbllx=315,bblly=30,bburx=752,bbury=474}
\epsfig{file=Gren_worldmods.ps, 
width=4.95cm,bbllx=5,bblly=30,bburx=443,bbury=474}
\end{center}
(c) \hskip 78mm (d)
\begin{center}
\epsfig{file=Gren_countcomp_noPS.ps, 
width=4.95cm,bbllx=315,bblly=30,bburx=752,bbury=474}
\epsfig{file=Gren_countrat.ps, 
width=4.95cm,bbllx=5,bblly=30,bburx=443,bbury=474}
\end{center}
\caption{The counts of lensed and unlensed field galaxies expected in a range of 
galaxy evolution scenarios (a) and world models (b). The wavelength 
dependence of the counts is compared in (c) and (d).}
\end{minipage}
\end{figure*} 

\vskip 0.5cm
{\centerline {4. A {\it FIRST} LENS SURVEY} }
\vskip 0.2cm

{\it FIRST} will provide a unique facility for submm-wave cosmology and will 
discover a host of new sources at large redshifts -- these will be valuable 
targets for future ground-based submm-wave interferometer arrays and 8-m 
telescopes operating in the optical/near-infrared waveband. Unaffected by 
atmospheric absorption, {\it FIRST} offers a unique 
capability to survey large areas of sky: large millimetre arrays [Ref.\,21,\,22] will 
provide sub-arcsecond resolution and sub-mJy sensitivities, but only in 
relatively small fields. {\it FIRST} would be very useful for observing 
molecular and fine-structure lines from known objects, which would otherwise 
be blocked by the atmosphere; however, the much larger antenna areas of 
ground-based submm telescopes would give them a significant advantage over
{\it FIRST} for conducting blank-field surveys [Ref.\,3].

Estimates of the number of lensed and unlensed galaxies that could be detected 
in a 0.2-year {\it FIRST} survey at 450\,$\mu$m are presented in Fig.\,3, based on a 
model of telescope performance [Ref.\,23] and the counts derived above. 
Estimates of the numbers of galaxies that could be detected in the extreme cases 
of a confusion-limited survey and an all-sky survey are listed in Table\,1; in the 
case of an all-sky survey the sensitivities of {\it Planck} [Ref.\,24] are assumed. 
All these estimates are derived under the assumptions of an evolving population 
of lensing galaxies, galaxy evolution model~A and an Einstein--de Sitter world 
model. The results can be scaled to match other scenarios using the counts in 
Fig.\,2, and the predicted number of lenses could be increased by a factor of 
about 5 without violating other observational constraints.

\begin{table*}[t]
\begin{minipage}{175mm}
\caption{$\sigma_{\rm cirrus}$ [Ref.\,25] and $\sigma_{\rm conf}$ are the cirrus 
and faint source confusion noise expected in a {\it FIRST} survey respectively, 
assuming that the mean galactic background intensity 
$I_{\rm 100\,\mu m}$ = 5\,MJy\,sr$^{-1}$ and $\sigma_{\rm conf}$ is the 
flux density of galaxies with a surface density of 0.03\,beam$^{-1}$.
$\sigma_{\rm sky}$ is the sensitivity of a {\it Planck} all-sky survey [Ref.\,24] and 
$\dot A_{\rm conf}$ is the rate of sky coverage that {\it FIRST} could achieve at a
5-$\sigma$ limit of $\sigma_{\rm conf}$. $N_{\rm conf}$ 
and $N_{\rm sky}$ are the number of 5-$\sigma$ sources expected in a 0.2-year 
confusion-limited survey and a {\it Planck} all-sky survey respectively.}
{\vskip 4pt}
\begin{tabular}{ p{1.2cm}p{1.2cm}p{1.6cm}p{1.2cm}p{3.4cm}p{3.9cm}p{1.2cm} }
$\lambda$ / & $\sigma_{\rm cirrus}$ / & $\sigma_{\rm conf}$ / & 
$\sigma_{\rm sky}$ / & $\dot A_{\rm conf} /$ & $N_{\rm conf}$ & $N_{\rm sky}$ \\
\end{tabular}
\begin{tabular}{
p{1.2cm}p{1.2cm}p{1.6cm}p{1.2cm}p{2.3cm}p{1.2cm}p{2.2cm}p{1.2cm}p{1.5cm} }  
$\mu$m&  mJy & mJy & mJy & deg$^{2}$\,day$^{-1}$ & lensed & 
unlensed & lensed & unlensed \\
\noalign{\vskip 4pt}
200  & 0.41 & 9.4 & 150 & 0.10 & 15 & $1.4\times10^4$ & 650 & $2.4\times10^5$ \\
450  & 0.51 & 14 & 100 &  0.22 & 16 & $6.6\times10^3$ & 560 & $7.1\times10^4$ \\
850  & 0.36 & 7.6 & 80 & 0.023 & 0.7 & 180 & 40 & $2.3\times10^3$ \\
\end{tabular}
\end{minipage}
\end{table*}

Between about 0.1 and 1\% of the sources detected in a {\it FIRST} survey are 
expected to be lensed by a foreground galaxy. At a wavelength of 850\,$\mu$m 
the surface density of detectable sources is expected to be rather small, despite
a relatively large ratio of lensed to unlensed galaxies. Hence, the only practical 
wavelengths for a lens survey are 200 and 450\,$\mu$m. The number of lensed 
galaxies expected at these wavelengths depends rather weakly on the limiting 
flux density, as shown in Fig.\,3. A 450-$\mu$m survey lasting 0.2\,years at a flux 
density limit of less than 50\,mJy would yield about 40 and 16 lenses at
signal to noise rations of 5- and 3-$\sigma$ respectively, amongst of order 
$10^4$ unlensed galaxies. This potential sample of systematically-selected 
lenses is comparable in size to the existing set of known lenses. Of 
course, the counts of unlensed galaxies in the submillimetre/far-infrared 
waveband would also be accurately determined in such a survey, leading to a 
more complete understanding of the evolution of the global star-formation rate 
at moderate redshifts.    

An all-sky {\it Planck} survey would detect of order 500 lensed galaxies, Table\,1. 
However, {\it Planck}'s 4-arcmin angular resolution is too coarse for rapid 
follow-up observations using ground-based telescopes, and the survey would
detect a very large number -- of order $10^5$ -- unlensed galaxies at 
wavelengths of 200 and 450\,$\mu$m. A total of only about 2300 candidate 
sources are expected at a wavelength of 850\,$\mu$m, and so {\it FIRST} could 
determine more accurate positions and flux densities for these galaxies in a 
short time -- several tens of hours. However, this would not be possible
for the much larger number galaxies detected at either 200 or 450\,$\mu$m. 
Despite this, the rewards from being able to sift through this large sample and 
select the lensed galaxies would be very considerable, and it is difficult to see 
how such a large sample of lenses could be compiled in any other waveband. 
Tens or hundreds of gravitational lenses could be detected in the submm 
waveband using {\it FIRST} and {\it Planck}, but how efficiently could these most 
valuable objects be separated from distant unlensed galaxies? 

\begin{figure}
\begin{center}
\epsfig{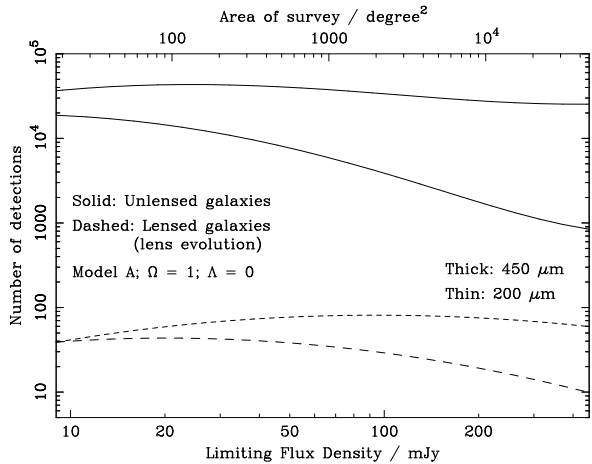}
\end{center}
\caption{The number of 3-$\sigma$ detections expected in 0.2-year {\it FIRST} 
surveys at 200 and 450\,$\mu$m.}
\end{figure}

\begin{figure}
\begin{center}
\epsfig{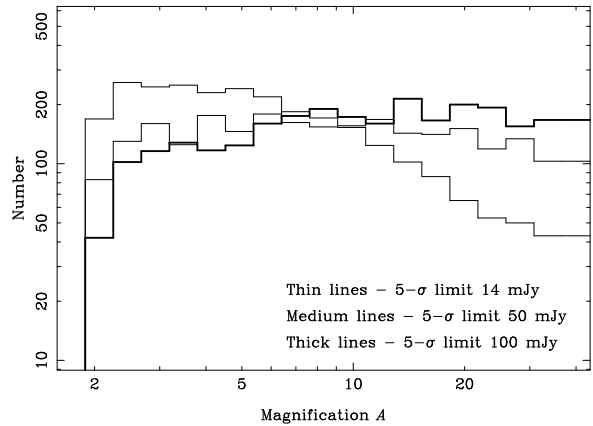}
\end{center}
\caption{The magnification distributions expected in 450-$\mu$m {\it FIRST} 
surveys at different depths. 
}
\end{figure}

The magnification distributions expected at three limiting flux densities at a 
wavelength of 450\,$\mu$m are shown in Fig.\,4. The distribution is tilted towards
larger magnifications at larger limiting flux densities; however, even in the 
deepest survey, most lensed galaxies would be expected to experience 
magnifications of order 4, and so a high-resolution image of a lens would be 
expected to show either several bright components, or to be significantly larger 
as compared with an unlensed galaxy at the same redshift. Hence, lensed and 
unlensed galaxies should be readily distinguishable in observations using large 
ground-based interferometer arrays. At a wavelength of 850\,$\mu$m these 
telescopes could detect a 10-mJy source at a signal-to-noise ratio of about 
60-$\sigma$\,beam$^{-1}$ in a 1-minute integration [Ref.\,21]. This sensitivity
and the 0.1-arcsec resolution of a large array should together be sufficient to 
allow any signs of lensed structures to be detected. Hence, a concerted and 
carefully-scheduled programme of follow-up observations using such an array 
could select the lensed sources in a sample of $10^4$ candidates in a few 
hundred hours. 

\begin{figure*}[t]
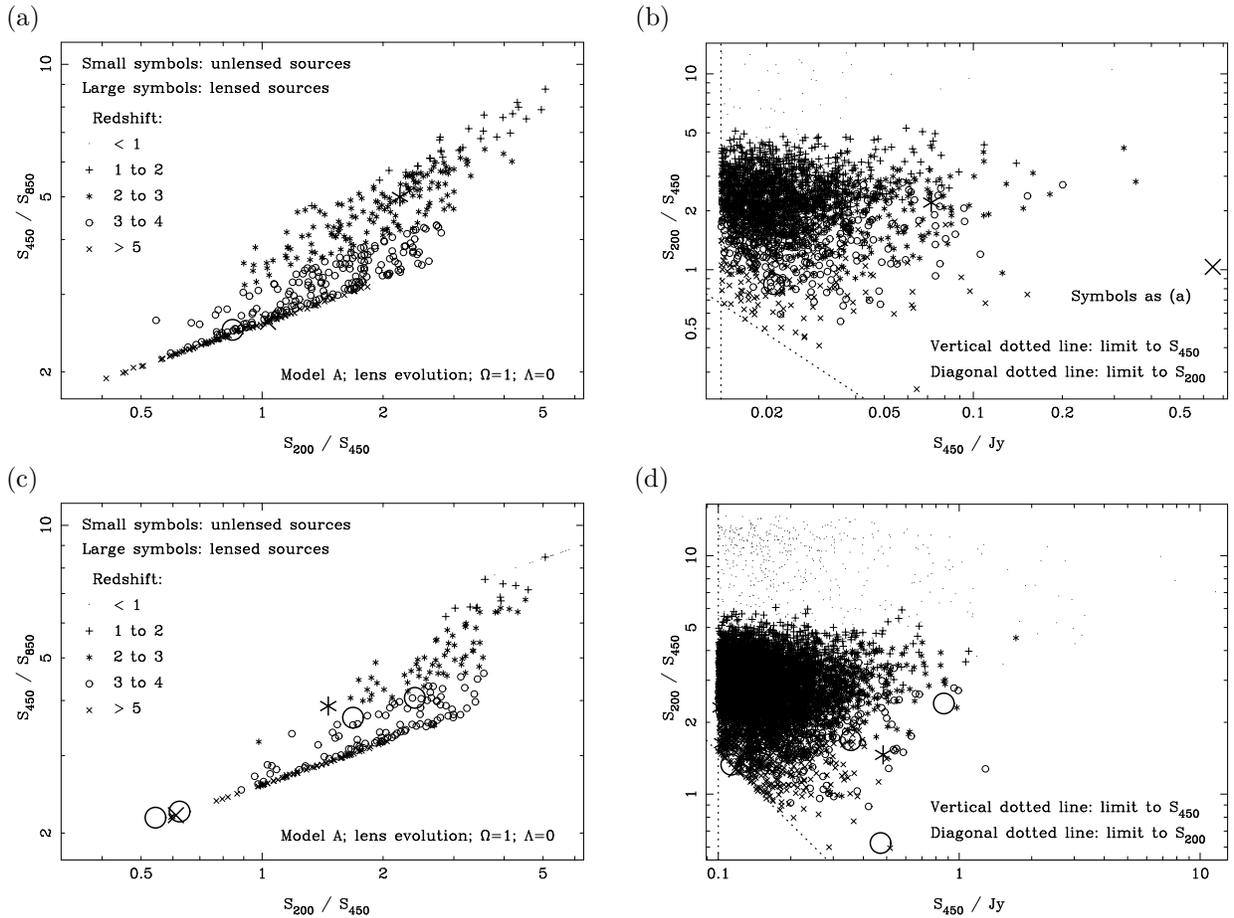

\begin{minipage}{170mm}
(a) \hskip 78mm (b)
\begin{center}
\epsfig{file=Gren_cc_CL.ps, 
width=4.95cm,bbllx=315,bblly=30,bburx=752,bbury=474}
\epsfig{file=Gren_cm_CL.ps, 
width=4.95cm,bbllx=5,bblly=30,bburx=443,bbury=474}
\end{center}
(c) \hskip 78mm (d)
\begin{center}
\epsfig{file=Gren_cc_WS.ps, 
width=4.95cm,bbllx=315,bblly=30,bburx=752,bbury=474}
\epsfig{file=Gren_cm_WS.ps, 
width=4.95cm,bbllx=5,bblly=30,bburx=443,bbury=474}
\end{center}
\caption{Colour--colour and colour--magnitude diagrams of typical samples of 
lensed and unlensed galaxies selected at 450\,$\mu$m in a confusion-limited [(a)
\& (b)] and all-sky [(c) \& (d)] survey. Careful selection of lensing candidates 
using these diagrams could increase the selection efficiency of a lens survey.}
\end{minipage}
\end{figure*}
 
Colour--magnitude and colour--colour diagrams for the simulated results of a 
confusion-limited and all-sky survey are presented in Fig.\,5. These results
suggest that such diagrams could be used to pre-select and order detected 
sources for ground-based imaging: lensed galaxies are found exclusively at 
large redshifts; and, if the dust temperature in star-forming galaxies is correlated 
with their luminosity, then lenses would be expected to have redder colours as 
compared with unlensed galaxies at the same flux densities. Careful
pre-selection of targets based on their colours in the submm waveband could 
reduce the size of the sample by a factor of a few, but high-resolution 
ground-based imaging of all the remaining candidates would still be required in
order to select lenses unequivocally.

The lensed galaxies detected in a {\it FIRST} survey would of course be subject 
to a detailed multi-waveband observing campaign. The properties of both the 
population and individual appearance of lenses in the sample would then be 
analysed in order to measure cosmological parameters and investigate the 
evolution of large-scale structure.  

\vskip 0.5cm
{\centerline {5. CLUSTERS} }
\vskip 0.2cm

Distant galaxies are also lensed by foreground clusters. In fact, the lensed 
images of distant galaxies are expected to be the brightest sources of
submm-wave radiation in the field of a cluster of galaxies on arcsecond-scales 
[Ref.\,11]. These images could not be resolved using {\it FIRST}, but their 
presence would be expected to increase the level of source confusion noise in 
the direction of the cluster -- by a factor of about 3 as compared with 
observations in the field [Ref.\,26,\,27]. The properties of this source confusion 
noise could be used to infer the form of evolution of faint galaxies in the submm 
waveband. Observations of a large sample of clusters would be required in order 
to investigate this effect; however, such a series of observations is already
planned to investigate another important feature of clusters in the submm
waveband -- the SZ effect [Ref.\,5,\,28].

The depth of a blank-field {\it FIRST} survey is limited by source confusion to 
flux densities of about $\sigma_{\rm conf}$ (Table\,1),  at which clusters with 
$y$-parameters greater than about $5\times10^{-5}$ and $5\times10^{-4}$ 
would be detectable at a signal to noise ratio of 5-$\sigma$ at wavelengths of 
850 and 450\,$\mu$m respectively. The surface density of detectable clusters is 
expected to be about 10 and 0.3\,deg$^{-2}$ at these wavelengths respectively 
[Ref.\,29]. A cluster would hence be detected between every 5 and 10\,days in a 
confusion-limited survey, a detection rate comparable to that expected for 
lensed field galaxies. 

Source confusion is expected to be much less significant in the mm
waveband [Ref.\,27,\,30], and so a {\it Planck} mm-wave all-sky survey has the 
potential to detect about $10^4$ clusters with $y > 8 \times 10^{-5}$. These 
clusters would provide valuable targets for pointed {\it FIRST} observations, to 
investigate both the effects of lensing discussed above and the submm-wave 
SZ effect [Ref.\,5]. 

\vskip 0.5cm
{\centerline {SUMMARY} }
\vskip 0.2cm

{\it FIRST}, especially in conjunction with the {\it Planck} mission, has the 
potential to detect and investigate the gravitationally-lensed images that are 
formed by both foreground galaxies and clusters. 
\vskip 0.1cm

i) The fraction of galaxy--galaxy lenses that could be detected in a {\it FIRST} 
survey depends on the properties of distant dusty galaxies and quasars, the 
observing wavelength and the flux density limit. It is expected to increase at 
longer wavelengths, and to be largest at the flux density at which the slope of 
the submm-wave counts begins to steepen at any particular wavelength. The 
surface density of detectable galaxies decreases at longer wavelengths, however, 
and is expected to be too small to allow a practical {\it FIRST} survey at a 
wavelength of 850\,$\mu$m. Several tens of lenses could be detected in surveys 
at wavelengths of 200 and 450\,$\mu$m; several tens of thousands of unlensed 
galaxies would also be detected, typically at moderate redshifts. First, 
submm-wave colours measured by {\it FIRST}, and secondly follow-up 
observations of suitable candidates using ground-based millimetre interferometer 
arrays and 8-m telescopes in the optical/near-infrared waveband would allow 
the lensed galaxies to be identified and studied further. The counts of unlensed 
galaxies could be used to constrain the evolution of the star-formation rate at 
moderate redshifts.
\vskip 0.1cm

ii) Several hundred lensed galaxies, and about $10^5$ unlensed galaxies, could 
be detected in a {\it Planck} all-sky survey at coarser resolution. Accurate flux 
densities and sub-arcminute positions for the most promising of these 
candidates could be obtained using {\it FIRST}, before similar follow-up 
observations. 
\vskip 0.1cm

iii) Several massive clusters could be detected due to their Sunyaev--Zel'dovich 
signal in a {\it FIRST} lens survey. Observations of a large sample of clusters, 
chosen from both those in pre-existing catalogues and the results of {\it FIRST} 
and {\it Planck} surveys, could be used to investigate the effects of lensing by 
clusters on the population of faint background galaxies in the submm waveband. 
These galaxies would otherwise be inaccessible to {\it FIRST}.

\vfill\eject
{\centerline {REFERENCES} }
\vskip 0.2cm

\begin{enumerate} 
\item{Franceschini A. et al., 1991, A\&AS, 89, 285}
\item{\vskip -0.2cm Blain A.\,W. \& Longair M.\,S., 1993, MNRAS, 264, 509}
\item{\vskip -0.2cm Blain A.\,W., 1996, in Shaver P. ed., {\it Science with large 
millimetre arrays}, Springer, Berlin, p.\,71}
\item{\vskip -0.2cm Blain A.\,W., 1996, MNRAS, 283, 1340}
\item{\vskip -0.2cm Rephaeli Y., 1995, ARA\&A, 33, 541}
\item{\vskip -0.2cm Schneider P., Ehlers J. \& Falco E.\,E., 1992, {\it Gravitational 
lenses}, Springer, Berlin}
\item{\vskip -0.2cm Kaiser N., 1992, ApJ, 388, 272}
\item{\vskip -0.2cmBorgeest U., Linde J.\,v. \& Refsdal S., 1991, A\&A, 251, L35}
\item{\vskip -0.2cm Knieb J.-P. et al., 1996, ApJ, 471, 643}
\item{\vskip -0.2cm Broadhurst T.\,J., Taylor A.\,N. \& Peacock J.\,A., 1995, ApJ, 
438, 49}
\item{\vskip -0.2cm Blain A.\,W., 1997, MNRAS, 290, 553}
\item{\vskip -0.2cm Peacock J.\,A., 1982, MNRAS, 199, 987}
\item{ \vskip -0.2cm Pei Y.\,C., 1995, ApJ, 404, 436}
\item{\vskip -0.2cm Press W.\,H. \& Schechter P., 1974, ApJ, 187, 245}
\item{\vskip -0.2cm Blain A.\,W. \& Longair M.\,S., 1996, MNRAS, 279, 847}
\item{\vskip -0.2cm Hewett P.\,C., Foltz C.\,B. \& Chaffee F.\,H., 1993, ApJ, 406, L43}
\item{\vskip -0.2cm Dunlop J.\,S. \& Peacock J.\,A., 1990, MNRAS, 247, 19}
\item{\vskip -0.2cm Oliver S.\,J., Rowan-Robinson M. \& Saunders~W., 1993, 
MNRAS, 256, 15P}
\item{\vskip -0.2cm Lilly S.\,J. et al., 1996, ApJ, 460, L1}
\item{\vskip -0.2cm Saunders W. et al., 1990, MNRAS, 242, 318}
\item{\vskip -0.2cm Brown R.\,L., 1996, in Bremer M.\,N. et al. eds., {\it Cold gas at 
high redshift}, Kluwer, Dordrecht, p.\,411}
\item{\vskip -0.2cm Downes D., 1996, in Shaver P. ed., {\it Science with large 
millimetre arrays}, Springer, Berlin, p.\,16}
\item{\vskip -0.2cm Beckwith S. et al., 1993, {\it FIRST}, SCI(93)6, ESA Paris}
\item{\vskip -0.2cm Bersanelli M. et al., 1996, {\it COBRAS/SAMBA}, SCI(96)3, ESA 
Paris}
\item{\vskip -0.2cm Helou G. \& Beichman C.\,A., 1990, in Kal-deich~B. ed.,
{\it From ground-based to space-borne sub-millimetre astronomy}. 
ESA vol. 314, p.\,117, ESA Paris}
\item{\vskip -0.2cm Blain A.\,W., 1997, in Holt S.\,S. \& Mundy J.\,G. eds., {\it Star 
formation: near and far}, AIP, Woodbury NY, in press}
\item{\vskip -0.2cm Blain A.\,W., MNRAS, submitted}
\item{\vskip -0.2cm Church S., this volume}
\item{\vskip -0.2cm De Luca A., D\'esert F.\,X. \& Puget J.-L., 1995, A\&A, 300, 335}
\item{\vskip -0.2cm Fischer M.\,L. \& Lange A.\,E., 1993, ApJ, 419, 433}

\end{enumerate} 

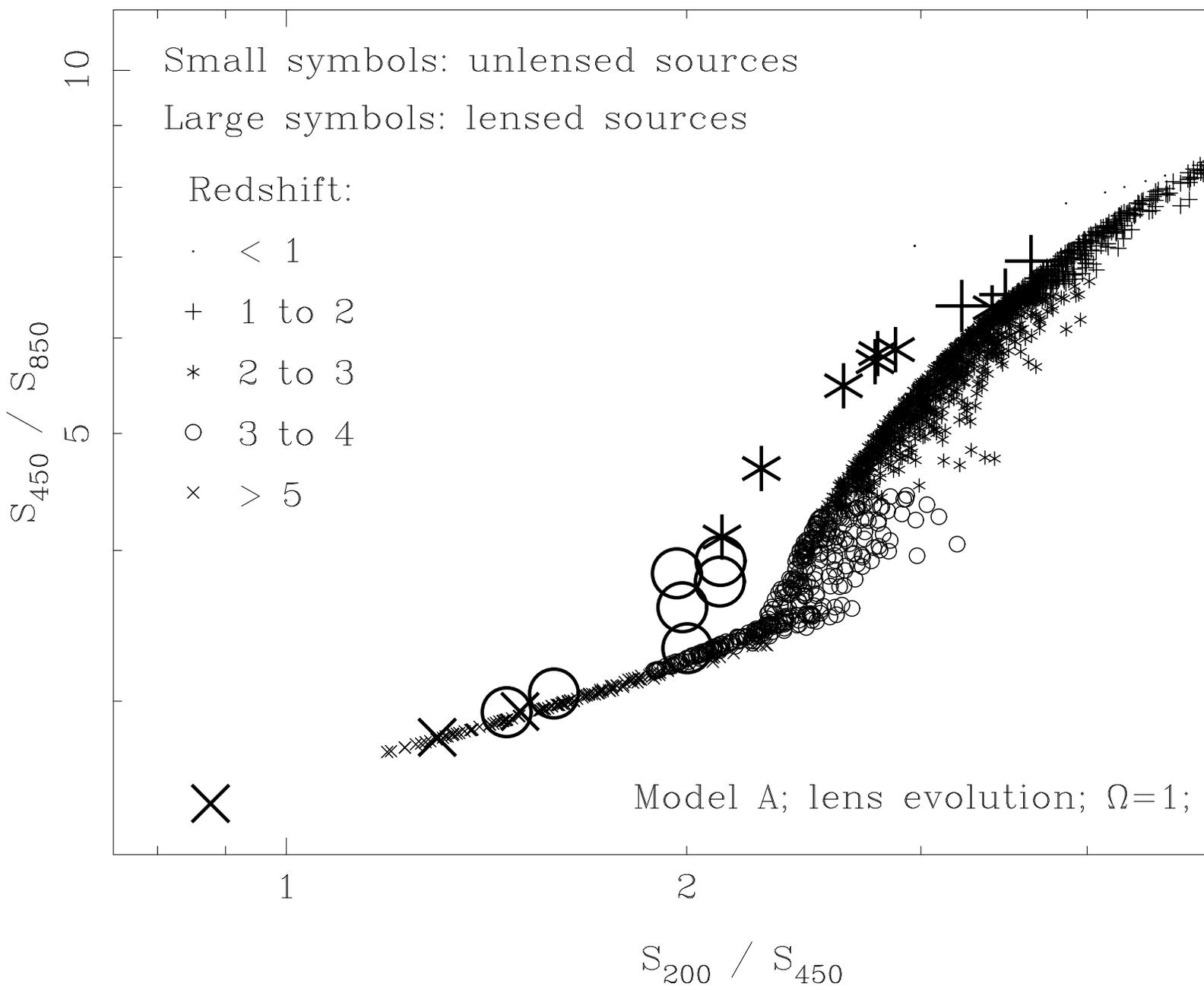
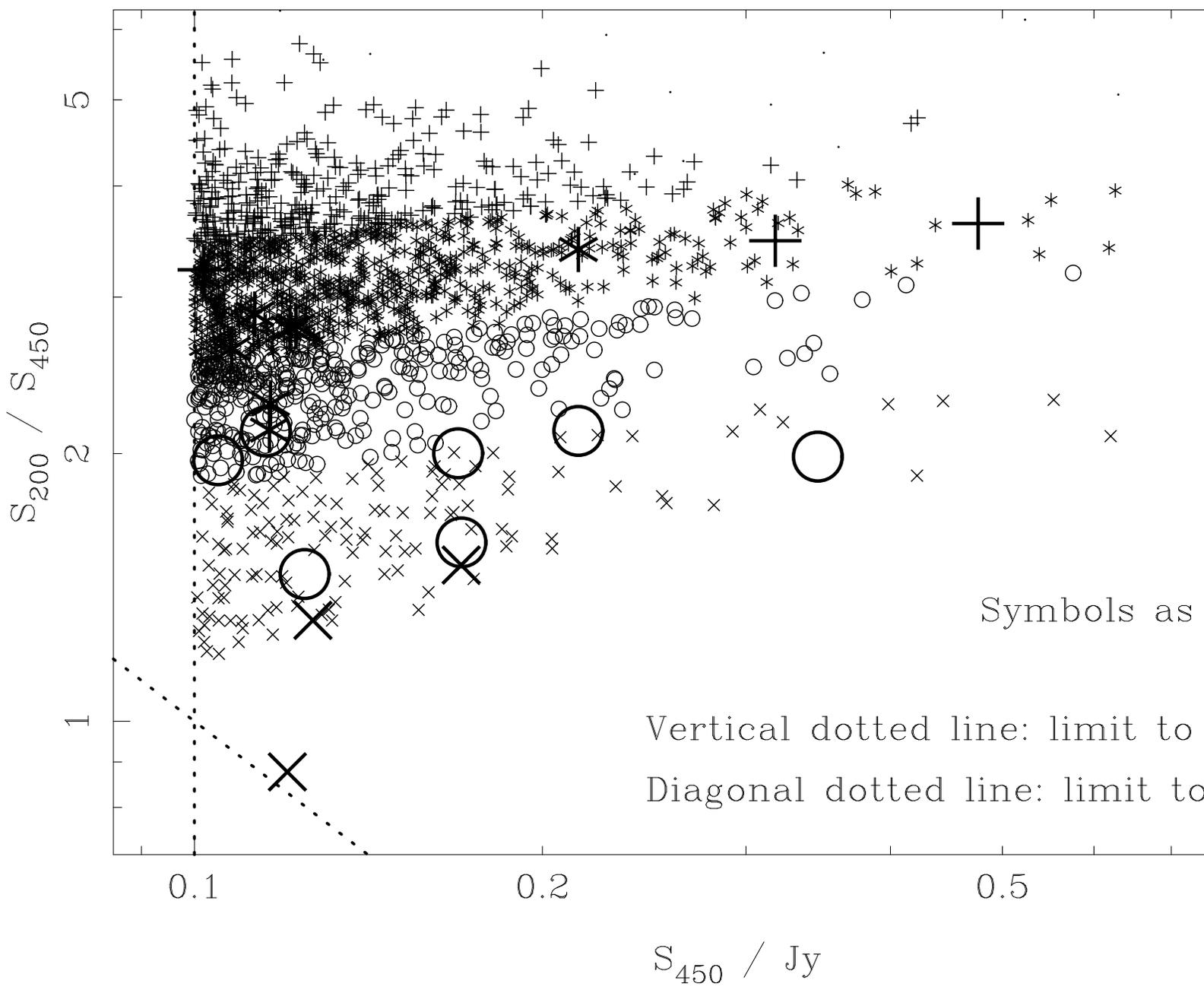
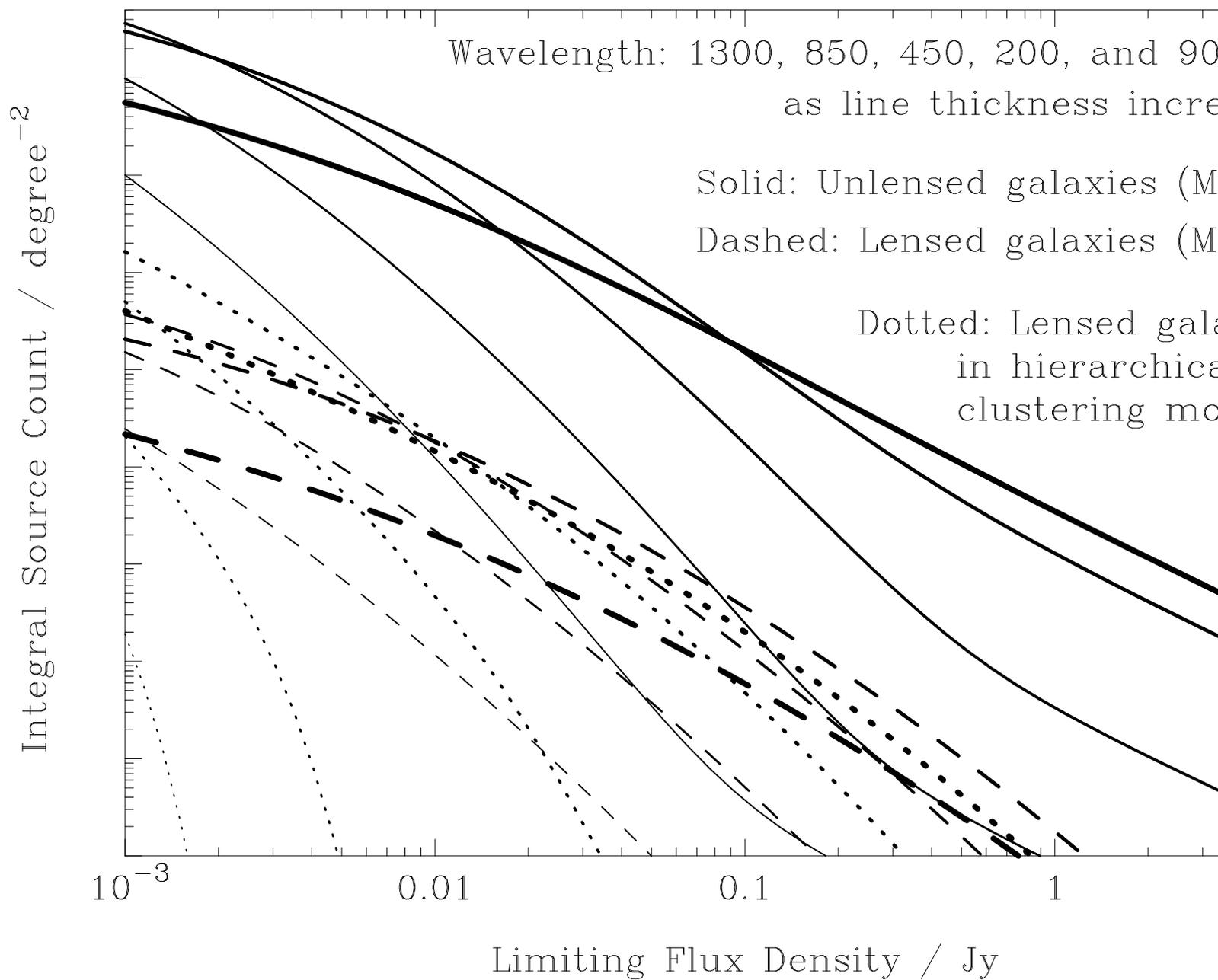
\end{document}